\documentclass{article}
\usepackage{authblk}
\usepackage[bottom=3cm, top=3cm, left=3cm, right=3cm]{geometry}
\usepackage{graphicx}
\usepackage{mathptmx}

\begin{document}

\title{Low-mass Dark Matter searches with EDELWEISS}

\author[1]{J. Gascon}
\author[2]{for the EDELWEISS Collaboration: E. Armengaud }
\author[1]{Q. Arnaud}
\author[1]{C. Augier }
\author[3]{A. Beno\^{i}t}
\author[4]{L. Bergé} 
\author[1]{J. Billard}
\author[4]{A. Broniatowski} 
\author[3]{P. Camus}
\author[1]{A. Cazes}
\author[4]{M. Chapellier} 
\author[1]{F. Charlieux}
\author[1]{M. De Jésus}
\author[4]{L. Dumoulin}
\author[5]{K. Eitel}
\author[1]{J.-B. Filippini}
\author[6]{D. Filosofov}
\author[4]{A. Giuliani}
\author[2]{M. Gros}
\author[7]{Y. Jin}
\author[1]{A. Juillard}
\author[8]{M. Kleifges}
\author[1]{H. Lattaud}
\author[4]{S. Marnieros} 
\author[1]{D. Misiak}
\author[2]{X.-F. Navick}
\author[2]{C. Nones}
\author[4]{E. Olivieri}
\author[4]{C. Oriol}
\author[9]{P. Pari}
\author[2]{B. Paul}
\author[4]{D. Poda}
\author[6]{S. Rozov}
\author[1]{T. Salagnac}
\author[1]{V. Sanglard}
\author[1]{L. Vagneron} 
\author[6]{E. Yakushev}
\author[5]{A. Zolotarova}

\affil[1]{{\small Univ Lyon, Universit\'e Lyon 1, CNRS/IN2P3, IP2I-Lyon, F-69622 Villeurbanne, France}}
\affil[2]{{\small IRFU, CEA, Université Paris-Saclay, F-91191 Gif-sur-Yvette, France}}
\affil[3]{{\small Institut N\'eel, CNRS/UJF, 25 rue des Martyrs, BP 166, 38042 Grenoble, France}}
\affil[4]{{\small Universit\'e Paris-Saclay, CNRS/IN2P3, IJCLab, 91405 Orsay, France}}
\affil[5]{{\small Karlsruher Institut f\"{u}r Technologie, Institut f\"{u}r Astroteilchenphysik, Postfach 3640, 76021 Karlsruhe, Germany}}
\affil[6]{{\small JINR, Laboratory of Nuclear Problems, Joliot-Curie 6, 141980 Dubna, Moscow Region, Russian Federation}}
\affil[7]{{\small C2N, CNRS, Universit\'e Paris-Sud, Universit\'e Paris-Saclay, 91120 Palaiseau, France}}
\affil[8]{{\small Karlsruher Institut f\"{u}r Technologie, Institut f\"{u}r Prozessdatenverarbeitung und Elektronik, Postfach 3640, 76021 Karlsruhe, Germany}}
\affil[9]{{\small IRAMIS, CEA, Universit\'e Paris-Saclay, F-91191 Gif-sur-Yvette, France}}


\maketitle

\begin{abstract}

The EDELWEISS collaboration searches for light Dark Matter (DM) particles using germanium detectors equipped with a charge and phonon signal readout. Using the Neganov-Trofimov-Luke effect, an rms resolution of 0.53 electron-hole pair was obtained on a massive (33.4~g) Ge detector operated underground at the Laboratoire Souterrain de Modane. This record sensitivity made possible a search for Dark Photon DM down to 1 eV/c$^2$ and to DM-electron interactions below 1 MeV/c$^2$. This demonstrates for the first time the high relevance of cryogenic Ge detectors in searches at low thresholds and is an important step of the development of Ge detectors with improved performance in the context of the EDELWEISS-SubGeV program.


\end{abstract}

\section{Introduction}

The absence of the discovery of a Weakly Interacting Massive Particle (WIMP)~\cite{rev} by large double-phase detectors~\cite{xenon}, coupled to the absence of the observation of beyond-the-Standard-Model particles at the LHC, has revived a keen interest for models where the Dark Matter (DM) particle would have so far escaped detection by having a mass below the GeV/c$^2$ range~\cite{lowmass} and a phenomenology departing from the standard WIMP scenario. 
The scattering of such particles -- present in our galactic halo -- on nuclei
could produce nuclear recoils with kinetic energies well below 1 keV. 
For DM particle masses below 100 MeV/c$^2$, the kinetic energy is most favourably  transferred to electrons,
either via a collision~\cite{dm-e} or an absorption~\cite{dp}.
In that context, direct DM search experiments are now developing  techniques to reduce the threshold for the 
detection of recoils (either a nucleus or an electron) to energies in the eV range, while keeping the 
very tight constraints on radioactive and instrumental backgrounds that are required for this very elusive signal.

The EDELWEISS experiment is in the process of adapting its cryogenic germanium 
detectors with phonon and ionization readout~\cite{edw-perf} to this new search domain. 
The aim of the EDELWEISS-SubGeV program is to search for DM particles with masses in the eV/c$^2$ to GeV/c$^2$ range using a 1-kg size array of germanium cryogenic detectors with particle identification capability, installed in its radiopure underground site at
the Laboratoire Souterrain de Modane (LSM).
This array would be ideally suited to detect DM scattering for wide range of particle masses: 
between 0.1 and 10 GeV/c$^2$ for scattering with nucleons, 
0.5 to 100 MeV/c$^2$ in the case of electrons, and between 0.7 and 10 eV/c$^2$ for the absorption of Dark Photons by electrons.
The choice of high-purity germanium as target takes advantage of the very low internal radioactivity of the material, its excellent properties that makes it an excellent medium for both charge collection and active volume for a massive bolometer, and its lower gap compared to silicon,
the target material mainly used by other projects aimed at these search domains~\cite{sensei,damic,cdms}.
The main limitation in these different searches is that the reduction of detection thresholds are obtained at the expense of particle identification capabilities. 
The EDELWEISS-SubGeV goal is to maintain these capabilities by achieving a resolution of $\sigma$ = 10 eV for the phonon channel and 20 eV$_{ee}$ (eV-electron-equivalent) for the ionization channel, while being able to ramp the charge-collection bias up to 100 V,
in order to exploit the Neganov-Trofimov-Luke (NTL) amplification of the phonon signal~\cite{luke}.
A 1~kg-scale array of germanium cryogenic detectors with such performance, operated in the low-background environment of the LSM,
could explore sub-GeV DM interaction with nuclear recoils with event-by-event rejection down to  a few 10$^{-43}$ cm$^2$, 
and exploit the NTL boost technique to explore MeV DM interaction with electrons (down to 10$^{-40}$ cm$^2$) and the absorption of eV-scale Dark Photon (down to $\kappa$=10$^{-15}$).
The required detector developments, in particular that of a cold HEMT preamplifier for the ionization readout~\cite{hemt}, 
are made in cooperation with the RICOCHET collaboration~\cite{ricochet}. 
Developments related to the  phonon channel improvement and to high-bias operation are described in the following.

\section{Recent results}

\begin{figure}[htbp]
\begin{center}
\includegraphics[width=0.49\textwidth,height=0.28\textheight]{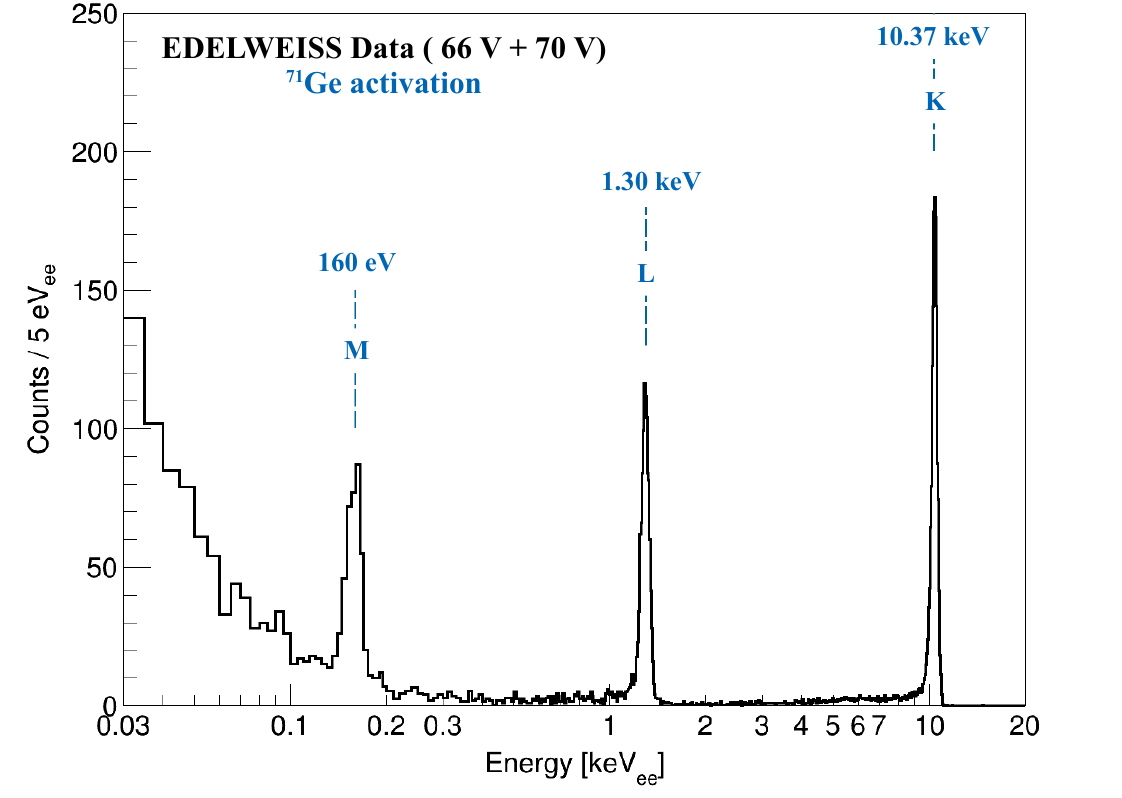}
\includegraphics[width=0.49\textwidth,height=0.29\textheight]{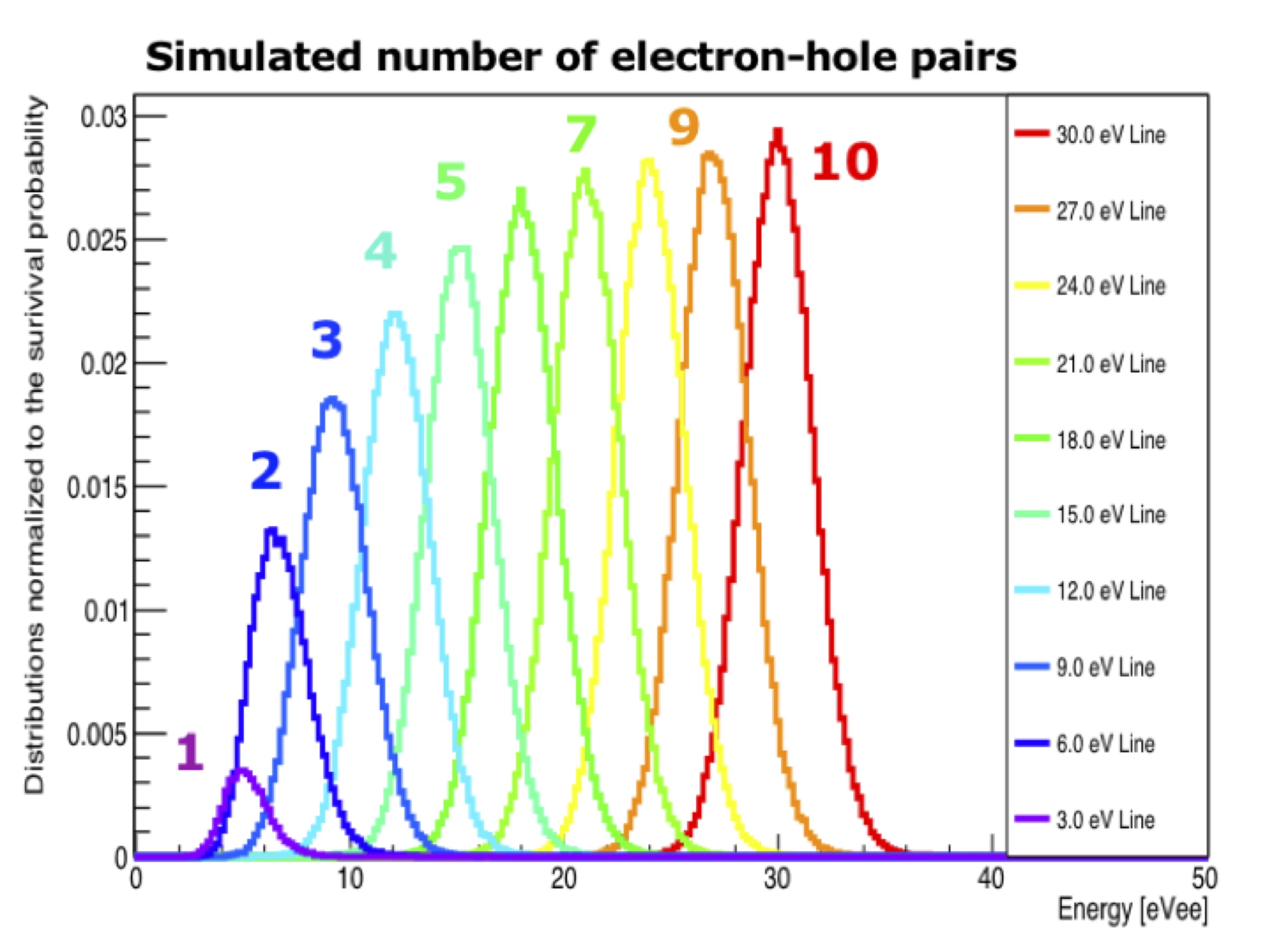}
\caption{Left panel:
Energy spectrum recorded in the phonon channel with a bias of 66 and 70 V
following the $^{71}$Ge activation of the detector. 
Right panel: 
Response of the detector to $N_{pair}$= 1 to 10 electron-hole pairs, obtained by inserting 
in the data streams appropriately scaled-down $^{71}$Ge K-line events. These simulated events
go through the same triggering, reconstruction and selection procedure used for the real data. \label{calib-resp}}
\end{center}
\end{figure}

In the course of this development process, the EDELWEISS collaboration has recently achieved a 17.7~eV 
phonon baseline resolution (rms) with a 33.4~g Ge bolometer operated above ground~\cite{edw-red20}. 
This detector provided the first sub-GeV constraints on DM-nucleon scattering.
To reach sub-electron-hole pair resolution, a similar detector was equipped with electrodes to take advantage of the NTL amplification.
It was operated at LSM in the ultra-low radioactivity environment of the EDELWEISS-III cryostat~\cite{edw-perf}, 
in order to perform a proper search of DM interaction with electrons.
The experiment, its results and their interpretation are described in more details in Ref.~\cite{edw-red30}.
The detector is a 33.4 g cylindrical high-purity Ge crystal ($\phi$ 20 $\times$ 20 mm), with 
electrodes lithographed on each of the two planar surfaces. 
The phonon sensor is  a Ge neutron-transmutation-doped (NTD), glued on the top of the crystal. 
The data from the phonon and ionization channels were digitized at a frequency of 100 kHz, filtered, averaged, and continuously stored on disk with a rate of 500 samples per second.

Prior to its installation in the cryostat, the detector was uniformly activated using a neutron AmBe source. 
The produced short-lived isotope $^{71}$Ge decays by electron capture in the K, L, and M shells.
The observed de-excitation lines at 10.37, 1.30, and 0.16 keV, respectively, are shown on the left panel of Fig.~\ref{calib-resp}.
The NTL amplification provided by applying a bias above 66~V is sufficient to clearly resolve the 160 eV line, 
with a width ($\sigma$ = 8 eV) only limited by the Fano factor. 
These lines are locally absorbed, and were used to probe the detailed response of the detector to an electron recoil signal 
uniformly distributed inside the detector volume. 
In particular, a sample of 858 events corresponding to K-line decays, recorded at the same bias used in the DM search (78~V), has been used to evaluate the efficiency of the entire data processing, triggering and selection process, as a function of deposited energy. These events were scaled to the desired energy and injected at random time in the data streams, and subjected to the same processing as the original data. The right panel of Fig.~\ref{calib-resp} shows the reconstructed energy distributions of the heat channel for events corresponding to 1 to 10 electron-hole pairs, {\em i.e.} 3 to 30 eV$_{ee}$ once the pair creation energy of 3 eV in cryogenic Ge is considered.
This event simulation procedure fully accounts for the actual noise conditions in the data streams selected for the DM search, and for all reconstruction and selection biases. 
The selection cuts are described in detail in Ref.~\cite{edw-red30}.
The efficiency for 3 eV$_{ee}$ events is 4\%. It rapidly increases to 22\% at 6 eV$_{ee}$ and plateaus close to 60\%. The baseline resolution for the search data is 1.58 eV$_{ee}$, corresponding to 0.53 electron-hole pair.

\begin{figure}[htbp]
\begin{center}
\includegraphics[width=0.54\textwidth,height=0.29\textheight]{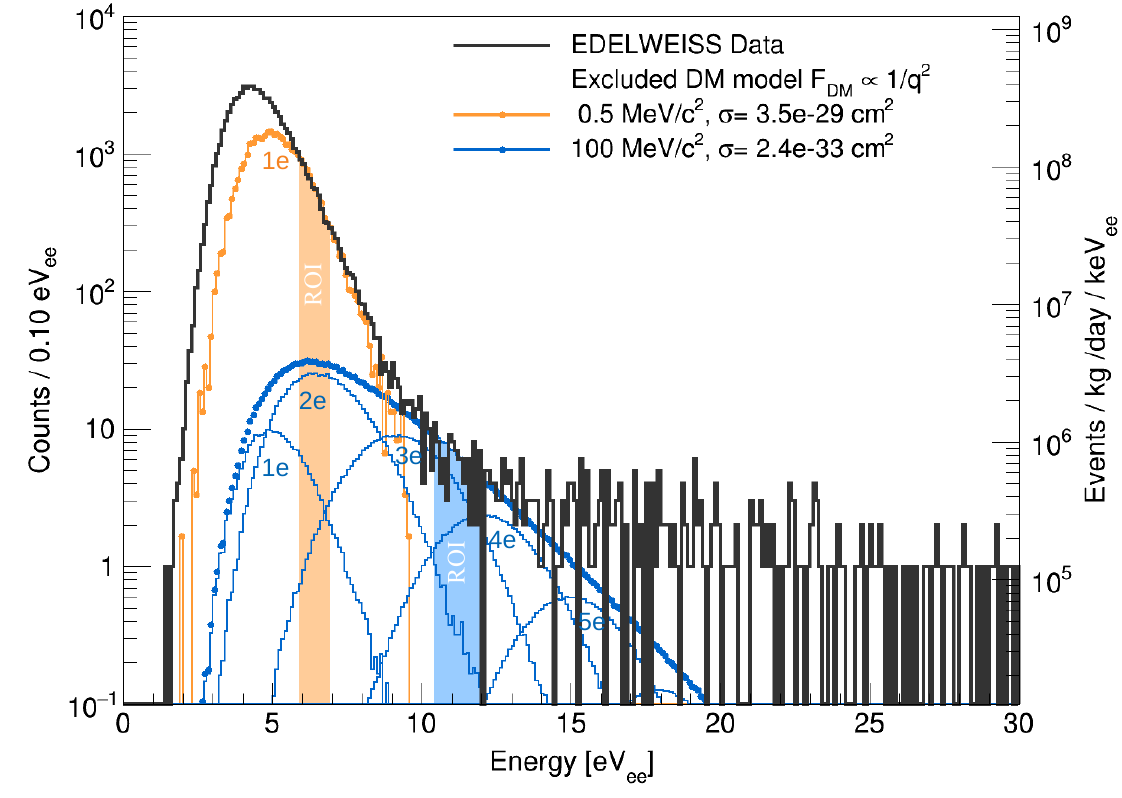}
\includegraphics[width=0.45\textwidth,height=0.30\textheight]{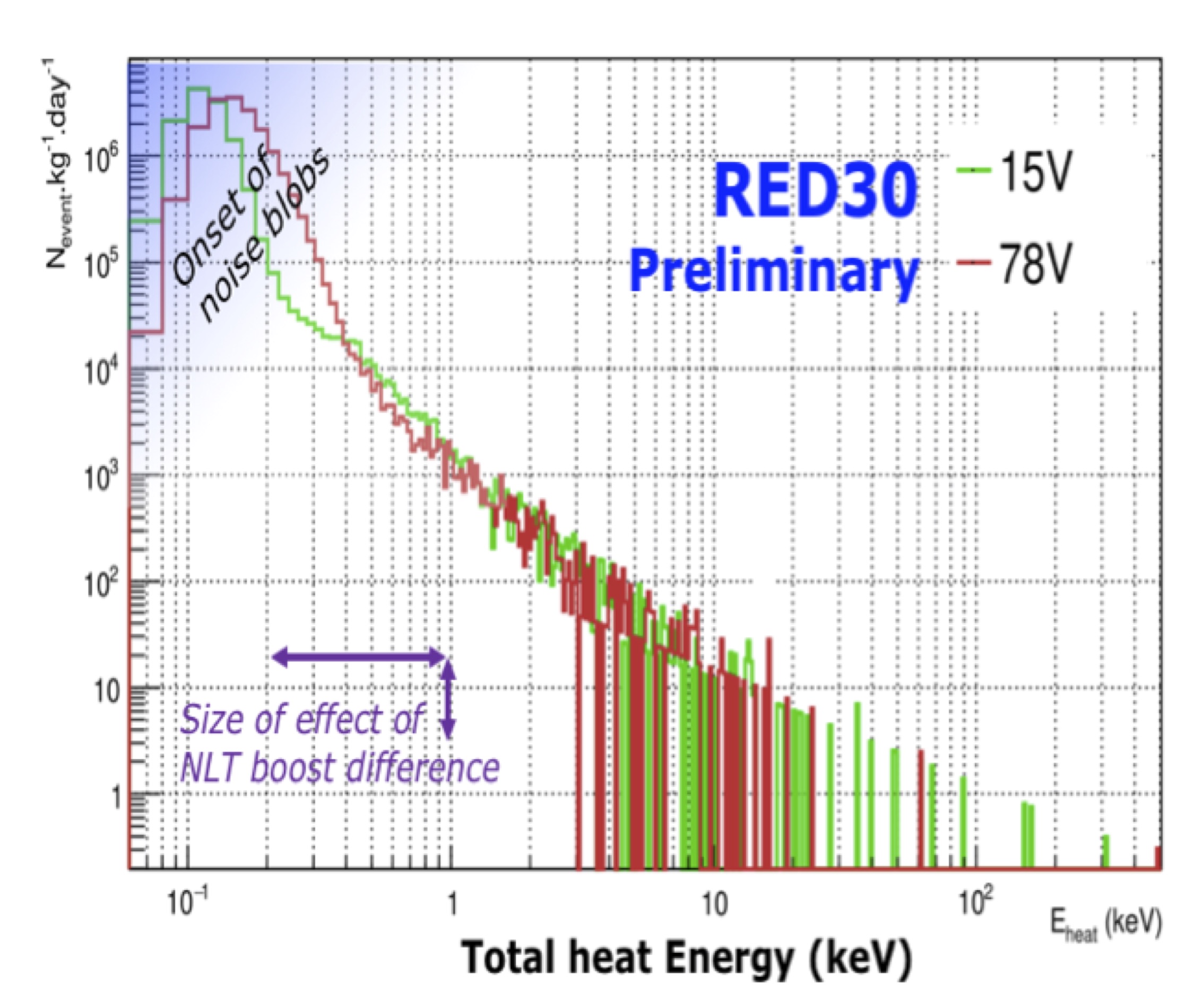}
\caption{Left:  Energy spectrum of the events selected for the DM search (black).
The thick blue (orange) histogram is the simulation of the signal excluded at 90\% C.L. for a DM particle with a mass of 10 (0.5) MeV/c$^2$, and $F_{DM} \propto 1/q^2$. The corresponding ROIs used to set the upper limits are shown as shaded intervals using the same colour code. The thin-line histograms of the same colour represent the individual contributions of 1 to 5 electron-hole pairs. Right: Total phonon energy spectra (corresponding to the energy in keV$_{ee}$ corrected for the NTL boost factor 1+V/3) recorded in the same detector at 15 and 78~V. In both panels, the data are not corrected for efficiencies. Instead, these are taken into account in the simulated signals. \label{fig2} }
\end{center}
\end{figure}

\begin{figure}[htbp]
\begin{center}
\includegraphics[width=0.329\textwidth,height=0.23\textheight]{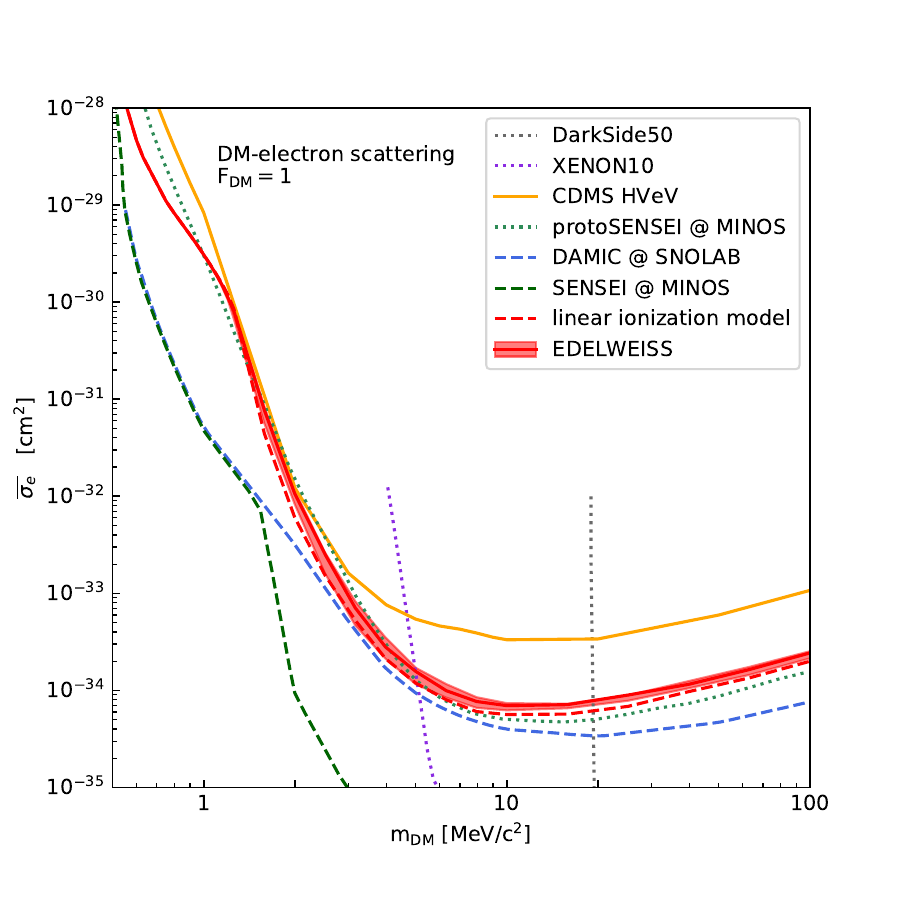}
\includegraphics[width=0.329\textwidth,height=0.23\textheight]{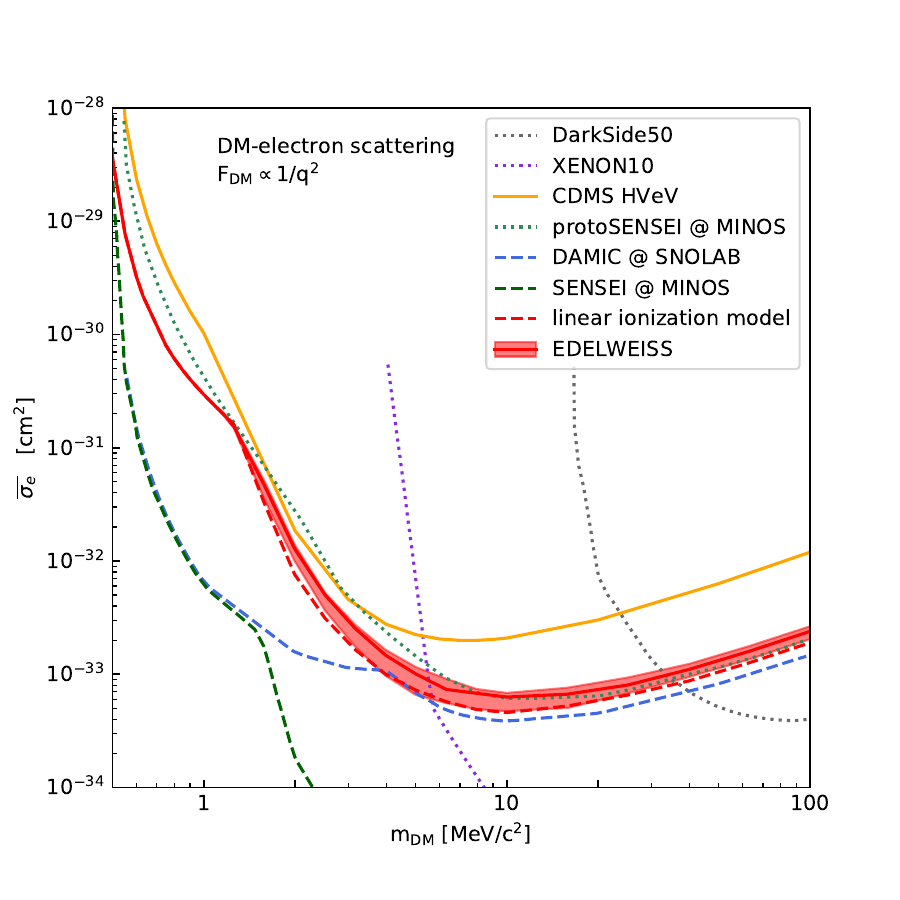}
\includegraphics[width=0.329\textwidth,height=0.23\textheight]{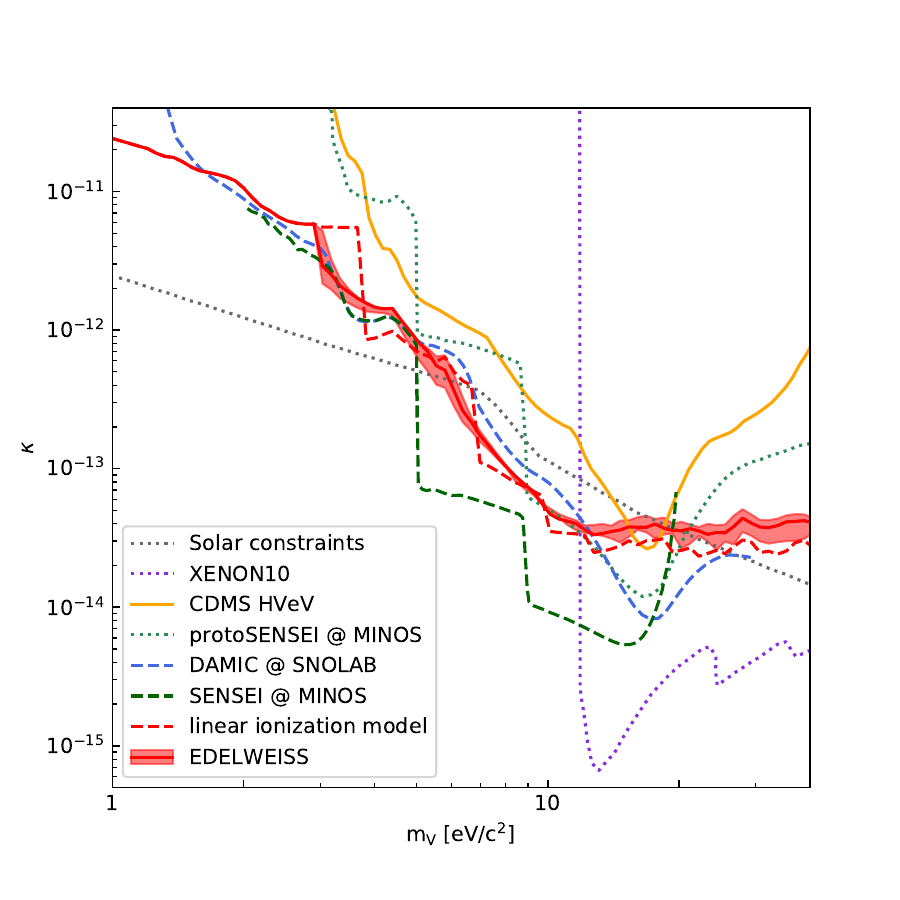}
\caption{
90\% C.L. upper limit on the cross section for the scattering of DM particles on electrons, assuming a heavy (left panel) or light (middle panel) mediator. Right panel: 90\% C.L. upper limit on the kinetic mixing $\kappa$ of a Dark Photon. The results from the present work are shown as the red line. The shaded red band and dotted red line represent alternative charge distribution models as described in Ref.~\cite{edw-red30}. Also shown are constraints from other direct detection experiments~\cite{dp,sensei,damic,cdms,bloch,darkside}, and solar constraints~\cite{solar}.
\label{fig3}}
\end{center}
\end{figure}

The search sample consists in 58 consecutive hours of data. 
The resulting spectrum is shown in the left panel of Fig.~\ref{fig2}.
All selection cuts and energy intervals used in the search have been determined using a non-blind sample of 21 and 9 hours taken immediately before and after the recording of the blind data, and in the same conditions.
Limits at 90\% C.L. on the interaction rate of DM particles with electrons were derived from the total number of events observed in selected energy ranges, as predicted by different models described below.
For each model, as a function of DM particle mass, the theoretical distributions were simulated using the same pulse simulation method used to determine the efficiency. 
The optimal energy range for the search was obtained using the non-blind sample, and the excluded cross sections are those that yield a number of counts exceeding the 90\% C.L. derived from Poisson statistics.
The considered models are the DM-electron scattering via a heavy or light mediator, corresponding to a form factor $F_{DM}=1$ or $1/q^2$, respectively~\cite{dm-e},
and the absorption of a Dark Photon via a kinematic mixing $\kappa$~\cite{dp}.
The resulting limits are shown in Fig.~\ref{fig3}.
The left panel of Fig.~\ref{fig2} shows examples of excluded spectra for two of the excluded models.

The present DM constraints extend to much smaller masses than searches based on noble gas detectors~\cite{bloch,darkside} and are competitive with those obtained with Si-based detectors~\cite{sensei,damic,cdms}. 

The better sensitivity of Ge compared to Si for a 1 eV/c$^2$ Dark Photon is due to the difference in gap energies. 
The present sensitivity for Dark Photons with masses below 3 eV/c$^2$ remains very competitive even with the
recent results obtained by SENSEI with a Si detector with a very small leakage current.
The EDELWEISS limit in this mass region derives from a 90\% C.L. upper bound of 4 Hz on the 
efficiency-corrected rate of single-pair events in the detector,
corresponding to a contribution to the leakage current of the detector of $<6.4\times 10^{-19}$ A.
Once scaled by the number of target electrons in the detector, 
the rate of single-charge events in EDELWEISS is 25 times larger than the limit obtained with Si by SENSEI~\cite{sensei}.
However, this higher background limit is offset by the more favorable gap energy in Ge.
In this respect, Ge is a more favorable target for low-mass Dark Photon searches. 

Further progress should come from the current work to improve the energy resolution, but an equally important issue is
to better understand the origin of the observed events. 
In addition to the improved constraints on Dark Matter models, 
the present improvement by an order of magnitude of the detection threshold for electron recoils compared to Ref.~\cite{edw-red20}
provides important insights in understanding the origin of the background limiting low-mass searches. 

The  resolution of the ionization channel (210 eV$_{ee}$) is not adequate to check if the events shown in the left
panel of Fig.~\ref{fig2} are indeed associated to the creation of electron-hole pairs in the detector.
However, the presence of charge can be tested by comparing the energy spectra recorded at different biases.
The total phonon energy $E_{tot}$ is the sum of the energy of the initial recoiling particle $E_0$, plus the contribution of NTL heating
due to the work done to drift the $N_{pair}$, $E_{NTL} = N_{pair}V$, where $V$ is the absolute value of the field applied to the detector. 
Using the average pair creation energy $\epsilon_{\gamma}$ = $E_0/N_{pair}$ (with $\epsilon_{\gamma}$ = 3 eV for electron recoils in Ge),
one gets that, for electron recoils and $V$ = 78~V, 
the total energy and the initial electron energies are related by 
\begin{equation}
E_{tot} = E_0 (1+V/\epsilon_{\gamma}) = 27E_0
\end{equation}
The right panel of Fig.~\ref{fig2} compares the total heat energy spectra recorded at bias of 15 and 78~V.
A cut on the ionization signal $E_{ion}<0$ is applied to remove the contribution of obvious electron recoil events.
This is found to be negligible compared to the observed rate for $E_{tot} < 10$~keV. 
Below $E_{tot} = 0.4$~keV, the energy resolution is not sufficient to unambiguously disentangle the contribution 
of the read-out noise, as the baseline resolution on $E_{tot}$ at 78~V is 10 to 20\% larger than at 15~V.
However, the $E_{tot}$ spectra between  0.4 and 10~keV are very similar at both bias values, 
and are not shifted by the factor 27 expected for electron recoils.
Therefore, most of the events above 15~eV$_{ee}$ on the left panel of  Fig.~\ref{fig2} are not electron recoils.
They correspond to a yet unidentified source of events that deposit energy in the detector without the creation 
of electron-hole pairs.
This type of "Heat-Only" population had already been reported by EDELWEISS~\cite{edw-3}, albeit at higher energy.
Below 15 eV$_{ee}$, the present detector resolution is not sufficient to unambiguously disentangle this population
from noise-triggered events or the possible contribution of leakage currents.
 
 The improvement by an order of magnitude of the detection threshold for electron recoils compared to Ref.~\cite{edw-red20}
 has thus not only  yielded interesting DM constraints, but also provided information that is important for the understanding of the origin of the background limiting low-mass DM searches. 

The work to improve both the phonon resolution of the detectors and the maximum bias at which they can be operated is ongoing.
A setup to characterize the energy response of cryogenic Ge detectors to single-photon excitations has been installed.
The first results have been presented at this conference~\cite{lattaud}. 
They confirm that Ge detector with NTL amplification do have the expected response assumed in the analysis of the LSM data.
Other developments concern the study and elimination of the Heat-Only event population, as it has the potential to be the
most important constraint in future searches in EDELWEISS~\cite{edw-prosp}. Systematic studies are under way.
Along these lines, another contribution to this conference~\cite{nbsi} describes a detector with a novel phonon sensor, 
consisting in a 100 nm thick high-impedance NbSi Transition Edge Sensor, lithographed as a 20 mm diameter spiral on
top of a 200 g Ge crystal. The use of this sensor eliminates the need of the glue used to fix the NTD sensor.
The NbSi film has sensitivity to athermal phonons, and opens the possibility to evaluate if Heat-Only events are purely thermal or not.
This sensor technique can be combined with NTL amplification. 
A resolution of 5 eV$_{ee}$ has already been achieved on a 200~g prototype operated at a bias of 66~V~\cite{nbsi},
and the physics analysis of the data is under way. 
The collaboration also investigates possible techniques to tag single-charge events in order to reject this background.

\section{Conclusion}

The objective of the EDELWEISS-SubGeV program is to develop a 1~kg-scale array of germanium cryogenic detectors to be operated at LSM to explore sub-GeV DM interaction with nuclear recoils with event-by-event rejection down to  a few 10$^{-43}$ cm$^2$, 
while keeping the ability to exploit NTL amplification in order to  explore MeV DM interactions with electrons (down to 10$^{-40}$ cm$^2$) and the absorption of eV-scale Dark Photon (down to $\kappa$=10$^{-15}$).
Significant recent progress on that path, with interesting physics results, is the achievement of a phonon resolution of
$\sigma=$17.3 eV on a 33.4~g detector, and of $\sigma$ = 0.53 electron-hole pair on a same-size detector biased at 78~V.
These developments yield to competitive DM search results, and in particular confirm the interest of germanium as
a low-gap target in searches where the signal is an electron recoil.
Further developments towards the EDELWEISS SubGeV goals are described in other contributions to this conference~\cite{hemt,lattaud,nbsi}.

\section{Acknowledgements}
The help of the technical staff of the Laboratoire Souterrain de Modane and the participant laboratories is gratefully acknowledged. The EDELWEISS project is supported in part by the German Helmholtz Alliance for Astroparticle Physics (HAP), by the French Agence Nationale pour la Recherche (ANR) and the LabEx Lyon Institute of Origins (ANR-10-LABX-0066) of the Universit\'e de Lyon within the program ``Investissements d'Avenir'' (ANR-11-IDEX-00007), by the P2IO LabEx (ANR-10-LABX-0038) in the framework ``Investissements d'Avenir'' (ANR-11-IDEX-0003-01) managed by the ANR (France), and the Russian Foundation for Basic Research (grant No. 18-02-00159). This project  has  received  funding  from  the  European Union’s Horizon 2020 research and innovation programme under the Marie Sk\l odowska-Curie Grant Agreement No. 838537. 
The datasets presented here are available from the corresponding author on reasonable request.

\end{document}